\documentclass{elsart}

\usepackage{graphicx}
\usepackage{mathrsfs}
\usepackage{amssymb}
\usepackage{amsmath}
\usepackage{graphics}
\usepackage{latexsym}
\usepackage{amsfonts}
\usepackage{color}
\usepackage{ctable}
\usepackage{verbatim}

\numberwithin{equation}{section}

\newcommand{\be}{\begin{eqnarray*}}
\newcommand{\ee}{\end{eqnarray*}}
\newcommand{\ben}{\begin{equation}}
\newcommand{\een}{\end{equation}}
\newcommand{\lb}[1]{\left[\begin{array}{#1}}
\newcommand{\rb}{\end{array}\right]}
\newcommand{\lp}[1]{\left(\begin{array}{#1}}
\newcommand{\rp}{\end{array}\right)}

\newcommand{\leftd}[1]{\left\{\begin{array}{#1}}
\newcommand{\rightd}{\end{array}\right.}

\begin{document}
\begin{frontmatter}

\title{Second law of thermodynamics in non-extensive systems}
\author[South]{J.P. Badiali}
\ead{jpbadiali@numericable.com},
\author[Oxford]{A. El Kaabouchi}
\ead{aek@ismans.fr}

\address[South]{Universit\'e Pierre et Marie Curie (Paris 6), 4 Place Jussieu, 75230 Paris 05, France} 
\address[Oxford]{Institut Sup\'erieur des Mat\'eriaux et M\'ecaniques Avanc\'es, 44, Avenue Bartholdi, 72000, Le Mans, France}

\begin{abstract} 

It exists a large class of systems for which the traditional notion of extensivity breaks down. From experimental examples we induce two general hypothesis concerning such systems. In the first the existence of an internal coordinate system in which extensivity works is assumed. The second hypothesis concerns the link between this internal coordinate system and the usual thermodynamic variables. This link is represented by an extra relation between two variables pertaining to the two descriptions; to be illustrative a scaling law has been introduced relating external and internal volumes. In addition, we use an axiomatic description based on the approach of the second law of thermodynamics proposed by E. Lieb and J. Yngvason (Physics Reports 310, 1999,1). We show that it exists an entropy function satisfying the monotony of the usual thermodynamic entropy. If a state results from the association of different states, the entropy is additive under these states. However, the entropy is a non-extensive function and we give its law of transformation under a change of the external volume. The entropy is based on some reference states, a change of these states leads to an affine transformation of the entropy. To conclude we can say that the main aspect of the second law of thermodynamics survives in the case of non-extensive systems.
\end{abstract}

\end{frontmatter}

\section{Introduction}

The thermodynamics is characterized by a very high degree of generality; many branches in physics are independent except that their results must be in accordance with thermodynamical predictions in the special case of equilibrium. The width of scope concerned by thermodynamics obliges us to express and discuss fundamental thermodynamic laws in terms of very general ingredients. Although statistical mechanics is frequently associated with thermodynamics we cannot identify these two fields of investigation. One goal of statistical mechanics is to calculate properties of a physical system starting from a description at a microscopic level, this requires to introduce a model for the system and to use a particular scheme of calculation. The main goal of thermodynamics is not to give the value of a given quantity but to establish general relationships between the properties of equilibrium states. Statistical mechanics adds something very useful to thermodynamics but it neither explains thermodynamics nor replaces it. In this paper we are strictly in a pure thermodynamic approach and we want to show that for non-extensive systems we may associate to any real transformation a non decreasing quantity.
 
The thermodynamics is based on two principles. The first appears as a conservation law and it is considered as granted. The main problem in a rigorous foundation of thermodynamics concerns the second law. E. Lieb and J. Yngvason (\cite{lieb}) have proposed to establish the second law on axioms that they considered as reasonable and completely intuitive. This fundamental work (noted LY hereafter) represents a step forward in comparison with the pioneer work of Giles (\cite{giles}). The LY approach is on the same line as the one proposed by Callen (\cite{callen}) for whom the entropy must be considered as a primary quantity, when expressed in terms of internal energy, volume and number of entities it gives the fundamental equation of thermodynamics. From this entropy representation we may derive other quantities like temperature or pressure. From this point of view the introduction of entropy does not require the concept of temperature that it is based on circular arguments (\cite{giles}) in traditional approaches (\cite{planck}) nor the concept of heath. The second law is expressed in the LY paper as an entropy principle saying that ``there is a real valued function on all states of all systems called entropy such that it verifies monotony, additivity and extensivity''.

In the LY paper, the extensivity is one of the corner stones from which the second law of thermodynamics can be established. However it exists a large class of experimental systems for which the extensivity must be abandoned. Due to the generality of the thermodynamics we think that the second law must also exist for such systems and our main goal is to show that it is so.

This paper is organized as follows. In Section $2$ we recall the definition of the extensivity in standard thermodynamic and we give examples of non-extensive systems; two hypothesis are introduced in order to caracterize non-extensive systems. In Section $3$ we summarize the main points of the LY work: we introduce the adiabatic accessibility, a pre-order relation, basic operations defined on state space and the axioms on which the approach of LY is based. In Section $4$ we elaborate the ingredients from which a thermodynamics of non-extensive systems may be proposed. In Section $5$ we investigate the relation between two couple of states and we will see that the pre-order relation leads to an algebraic inequality. In Section $6$ we first introduce the concept of adiabatic equivalence in standard thermodynamics from which some ingredients used in LY become more evident and then the adiabatic equivalence is derived for non-extensive systems. Section $7$ is devoted to the definition of the entropy : monotony and additivity are deduced but the extensivity is lost and a scaling law showing how the entropy is changed by a dilation process is given. Conclusions are given in the last Section.

\section{Extensivity and non-extensivity} 
A physical system $S$ can be observed in different stationary states that we can identify from an experimental point of view and to which we may associate a given label. In LY, the existence of a coordinates system is not needed to refer these states however to make a connection with the standard thermodynamic description it might be helpful to introduce a thermodynamic coordinate system $(TCS)$. For a mono-phasic mono-component system at equilibrium that we consider as an illustrative example hereafter, it is well known (\cite{callen}) that the $TCS$ is formed of three independent coordinates. The external volume $V$ in which the system is contained, it is the relevant mechanical parameter. The internal energy $U$ associated with a given reference state, in presence of impermeable adiabatic walls the change in $U$ is equal to the mechanical work performed on the system by external devices. Finally, we have to specify the number $N$ of particles in the system. The values of $V,U$ and $N$ can be changed independently, arbitrarily and whatever the precise nature of investigated system. Of course the choice of $(V,U,N)$ is not unique it is just the most used.

\subsection{Extensive systems} 
Standard thermodynamics is based on the following ingredients : A state ${X = (V, U, N)}$ can be also caracterized by the coordinates ${(V,\frac{U}{V}, \frac{N}{V})}$ where the 
quantities ${\frac{U}{V}}$ and ${\frac{N}{V}}$ are assumed independents of $V$ and are called the intensive variables. A state defined by ${X(t)=(tV,tU,tN)}$ is assumed to exist and we have    
${(tV,\frac{tU}{tV},\frac{tN}{tV})=(tV,\frac{U}{V},\frac{N}{V})}$. This means that changing $t$ we may generate an infinite number of systems having the same values for the intensive parameters but different extensions in space. The passage from $X$ to $X(t)$ is called a dilation process and in this case we can note $X(t) = tX$. This first definition of extensivity associated with a state $X$ is extended in two directions. First, it is assumed that in a state $X$ it exists extensive properties ${Q(X)=Q(V, U, N)}$ for which we have ${Q(X)= Vq(x)}$ where ${q(x)}$ is a given function of intensive parameters noted $x$. If we perform a dilation of $t$ we will have

\begin{equation}
Q(X(t)) = Q(tX)= tVq(x) = tQ(X)
\label{qext}
\end{equation}

showing that $Q(X)$ obeys to a homogeneous first order function of the extensive parameter. It is assumed that (\ref{qext}) is verified in the case of the entropy ((\cite{callen}) page 28). 
The second generalization of extensivity consists in assuming that if the transformation of a state $A$ in a state $B$ is possible then the transformation may persist for any amount of matter leading to 

\begin{equation}
\forall t > 0, (A\rightarrow B)\Rightarrow((A(t)\rightarrow B(t))~{\rm or}~(tA \rightarrow tB))
\label{defext}
\end{equation}

\subsection{Non-extensive systems}
\subsubsection{Examples of non-extensive systems}
In a large class of systems (polymers, colloidal solutions, natural see water, materials in presence of fractures,  ...) we observe the formation of structures leading to the existence of power laws between relevant quantities (see for instance (\cite{pgg})). Frequently this is the signature of the existence of a fractal dimensions (\cite{mandelbrot}), (\cite{feder}). As an example we consider a system in which the formation of aggregates is observed in presence of a supporting fluid considered as having fixed properties. A such system behaves as an effective one component system: $U$ is now the energy associated with aggregates, $N$ the number of particles forming the aggregates and $V$ is the volume of the box in which the system is contained. A typical length $l_{0}$ can be associated with $V$ that we write $V = f_{0}l_{0}^{3}$ in which $f_{0}$ is a form factor. Inside $V$, depending of the nature and concentrations of active material, temperature of the thermostat with which it is in contact, ... there is an internal volume $v$ representing the volume occupied by the aggregates. We may define $v$ as 

\begin{equation}
v = V\left(\frac{l}{l_{0}}\right)^{D} = \left(f_{0}^{\frac{D}{3}} l^{D}\right) V^{1 - \frac{D}{3}} \approx V^{1 - \frac{D}{3}}
\label{smallv}
\end{equation} 

in which $l$ is a characteristic size of the aggregate and $D$ a fractal dimension. The relation (\ref{smallv}) suggests the existence of a general power law $v \propto V^{d}$ relating the internal volume $v$ and the macroscopic one $V$. If, for instance we know $N$ the number of monomers introduced in the volume $V$ we can also write $ v \propto N^{d}$ and if we characterize the volume by its radius of gyration we get the well know Flory relation (\cite{pgg}). Any property related to $v$ will exhibit a specific power law. For instance if the interaction between active species corresponds to a radial potential $v(r) \propto (\frac{1}{r})^{\alpha}$ in which $r$ is the mean distance between the particles that we may estimate to be in average $(\frac{v}{N})^{\frac{1}{3}}$, the total energy will be $U \propto N (\frac{N}{v})^{\frac{\alpha}{3}} = N (\frac{N}{V^{d}})^{\frac{\alpha}{3}}$ showing that the total energy $U$ does not exhibit a linear dependence through $N$ or $V$. In parallel to experiments it exists numerical simulations showing to the formation of fractal agregates (see for instance (\cite{dla1}, \cite{dla2})).

In principle the fractal dimension may depend on the state of the system. When real materials are submitted to strong deformations it may appear fractures giving rise to fractal structures for which the fractal dimension may change with the magnitude of the external stress (\cite{deform}). It has also been observed that the fractal dimension of an electrode entering in the constitution of an electric battery changes after a discharge; from this result it has been suggested for the first time that the entropy may contain a fractal dimension (\cite{lemehaute}).

\subsubsection{Hypothesis concerning the non-extensive systems}
From the state $X =(V, U, N)$ of a non-extensive system we can create a state $tX$ of coordinates $(tV, tU, tN)$. However in a dilation process the quantity to be invariant is $\frac{U}{v}$ but not $\frac{U}{V} \approx \frac{U}{v^{\frac{1}{d(x)}}}$. Thus, in opposite to the case of extensive systems the transformation of a state $X$ into $tX$ cannot be considered as a dilation. The dilation by $t$ of a non-extensive system will be noted $X(t)$.

Instead of the $TCS$ defined by $(V, U, N)$ it seems natural to consider an internal coordinate system $ICS$ in which a state noted $\bar{X}$ is referred by $(v, U, N)$ or by $\bar{X} = (v,\frac{U}{v}, \frac{N}{v})$. Since both $U$ and $N$ are localized in $v$, $\frac{U}{v}$ is the density of energy in the aggregates and $\frac{N}{v}$ the density of particles inside the aggregates. The internal volume $v$ is an experimental quantity like $V$;
for instance in the case of polymer physics (\cite{pgg}) the gyration radius can be determined by neutron or ligth scattering. However $v$ does not exist for any system and since it depends on a molecular description $v$ is not a thermodynamic quantity. Indeed, we cannot act on $v$ as in the case of the external volume $V$ that is directly under our control. This can be illustrated by considering the Joule experiment (see (\cite{planck}) p.44), in which we can change $U$ maintaining $V$ constant but $v$ may change because in the Joule experiment the temperature is changed.

For a large class of systems we assume that $\frac{U}{v}$ and $\frac{N}{v}$ are independent of $v$ and therefore that a traditional dilation process exists in the $ICS$; for a 
dilation of $a$ we note $ \bar{X}(a) = a \bar{X}$ and to $ \bar{X}(a)$ of coordinates $(av, aU, aN)$ or $(av,\frac{U}{v}, \frac{N}{v})$. The existence of states having the same internal intensive parameters but differing only by their extension is in agreement with some experimental facts. For instance, in the case of real systems in natural sea water we observe a given distribution of aggregates (see for instance (\cite{wilkinson}), (\cite{logan})) from which we may determine  a fractal dimension. The traditional box counting (\cite{feder}) is replaced by an experimental analysis on the aggregates forming the distribution, they are assumed to have the same internal intensive properties and then the same fractal dimension the only difference between them is their extensive property (volume, maximum length, ...). In a polymeric solution we assume that there is a distribution of polymers having the same intensive properties but different lenghts. 
From these arguments we introduce our first hypothesis concerning the non-extensive systems:

$H_{1}$ - For any non-extensive system in parallel to the $TCS$ it exits a $ICS$ in which the system can be considered as extensive.

This means that in $ICS$ we have the properties already developed in the subsection $2.A$ devoted to extensive system. In particular, it exists some properties $\bar{Q}(\bar{X})$ for which we have $\bar{Q}(\bar{X}) = v \bar{q}(\frac{U}{v}, \frac{N}{v})$ and   

\begin{equation}
\bar{Q}(\bar{X}(a))=\bar{Q}(a\bar{X})=av\bar{q}\left(\frac{U}{v},\frac{N}{v}\right) =a\bar{Q}(\bar{X})
\label{pnonext}
\end{equation} 

To link (\ref{pnonext}) with thermodynamics we must express $a$ in terms of thermodynamic quantities. This can be done by introducing an extra relation between $v$ and $V$. Our second hypothesis is the following

$H_{2}$ - Between the extensive properties in $TCS$ and $ICS$ it exists an extra relation that we can know from experiments or theoretical predictions.

Previously $v$ and $V$ have been considered as the two extensive quantities and we have introduced a power law $v = A V^{d(x)}$ in which $A$ and $d(x)$ are two positive quantities depending on the state of the system via its intensive properties. If the external volume $V$ is expanded linearly $V \rightarrow tV$ as a consequence of $H_{2}$ the internal volume becomes 
$v(t) = t^{d(x)}v$. To realize a dilation in this case we have to change $U$ in $t^{d(x)}U$ and similarly $N$ in $t^{d(x)}N$. Such a dilated noted $X(t)$ is different from $tX$ defined above. By combining $H_{1}$ and $H_{2}$ in a dilation process of a given state we have

\begin{equation}
V\rightarrow tV \Rightarrow X \rightarrow X(t)
\label{vt}
\end{equation}

\begin{equation}
\bar{X} \rightarrow  \bar{X}(a)= a \bar{X} = t^{d(x)}\bar{X}
\label{xt}
\end{equation}

For a scalar extensive property $Q$ we can complete (\ref{pnonext}) according to 

\begin{equation}
\bar{Q}(a\bar{X})=a Q(\bar{X})=t^{d(x)}\bar{Q}(\bar{X})=t^{d(x)}Q(X)=Q(X(t)).
\label{qnonext}
\end{equation}

Since the value of a property is independent on the choice of the coordinate system.
    
\section{Axiomatic approach of the second law of thermodynamics in the case of extensive systems.}
In this Section we briefly summarize the work of Lieb and Yngvason (\cite{lieb}). 
\subsection{States and transformation of states}
To found thermodynamics avoiding the introduction of heath as a primarily quantity we focus first on mechanical processes as in the route followed by Caratheodory. 
In the LY work the corner stone is the existence of a pre-order relation between two states $X$ and $Y$, noted $X \prec Y$, it means that the passage from $X$ to $Y$ can be realized just by observing a change in a mechanical device external to $S$, all other ingredients appearing in the transformation are in the same state at the initial and final level of the transformation. The concept of adiabatic accessibility characterized by $\prec$ is more general that an adiabatic transformation in standard thermodynamics however it has been shown in LY that to the adiabatic accessibility we may associate a standard adiabatic transformation: hereafter to be short we will consider these two terms as equivalent.  The adiabatic accessibility does not assume that the passage from one state to another is done slowly or in a quasi equilibrium manner. If in addition to $X \prec Y$ we have also $Y \prec X$ we say that $X$ and $Y$ are adiabatically equivalent and we note $X \approx Y$. When we have $X \prec Y$ but not $Y \prec X$ we note $X \prec \prec Y$. The relation $X \prec Y$ is a relation between states but not between numbers, the beautiful result obtained by LY is that the axioms associated with $\prec$ are sufficient to introduce a real valued function verifying the main properties of the entropy.
  
An important operation on $\Gamma$ consists in the association of two states $X$ and $Y$ in order to create a couple noted $(X,Y)$ that is an element of a space-product noted $\Gamma \times \Gamma$. To obtain $(X,Y)$ we duplicate $S$, one part is putted in a state $X$ and enclosed in a box having adiabatic walls concerning the second part we perform the same procedure but with the system taken in a state $Y$. If the two boxes with their walls are putted together side to side we have one realization of $(X,Y)$ but this juxtaposition of states is just a first step that we can realize in any case. In a second step many processes are possible and $(X,Y)$ can be adiabatically transformed into $(X',Y')$.

Hereafter to the state $X$ we associate $X(t)$ also noted $tX$ for $t > 0$. Keeping $t>0$, it is useful to give a meaning to $- tX$. By definition, when $-tX$ appears on one side of a transformation it has been understood that it can be put as $tX $ on the other side of the transformation, thus, for example, 

\begin{equation}
(X) \prec (Y, -tZ)  \Rightarrow (X,tZ) \prec (Y). 
\label{sign}
\end{equation}

Beyond the formation of couple we also introduce the formation of n-uple for which we assume permutativity and associativity of the states. This means, for example, that we may write 

\begin{equation}
(X, Y, Z) \equiv (X, Z, Y) \equiv ((X,Y),Z) \equiv (X,(Y,Z)) ...
\label{asso}
\end{equation}

\subsection{Axioms and hypothesis}
The first two axioms $A_{1}$ and $A_{2}$ used in LY are traditional for any order relation. $A_{1}$ is associated with the reflectivity $(X \approx X)$ and $A_{2}$ is concerned by the transitivity, it means that if we have $X \prec Y$ and $Y \prec Z$ then we have $X \prec Z$. The axiom of consistency, $A_{3}$, says that if we have $X \prec X'$ and $Y \prec Y'$ then there is an adiabatic transformation for which we have $(X, Y) \prec (X', Y')$. We may illustrate this axiom using the example given in the previous subsection. We can create $(X,Y)$ using adiabatic boxes, if we maintain all the walls and transform separately the contents of each box there is no doubt that we will obtain $(X', Y')$. However, if we have $(X, Y) \prec (X', Y')$ we can not conclude that $X \prec X'$ and $Y \prec Y'$ in general.

With the axiom of scaling invariance, $A_{4}$, it is assumed that if we have $X \prec Y$ then for any $t > 0$ we will have $tX \prec tY$. 
The axiom $A_{5}$ is related to splitting and recombinaison of one state, it claims that for any $0 < t <1 $ we have the adiabatic equivalence $X \approx (tX,(1-t)X)$. This can be illustrated by considering the previous two boxes separated by an adiabatic wall. If one box is filled by a system in a state $tX$ and the second one by a state to $(1-t)X$ it seems obvious that if take off the adiabatic wall separating the two system we will have $(tX,(1-t)X) \prec X$. Now if we start with a box containing $X$ we may introduce a wall separating $X$ into $tX$ and $(1-t)X$ this can be done adiabatically and we have $(X \prec (tX,(1-t)X)$. Thus this axiom seems obvious. The last axiom, $A_{6}$, says that if $(X, \epsilon Z_{0}) \prec (Y, \epsilon Z_{1})$ holds for a sequ
ence of $\epsilon \to 0$ and some states $Z_{0}$ and $Z_{1}$ then we have $X \prec Y$. From this axiom we see that it is impossible to increase the set of accessible states with an infinitesimal grain of dust.

To these quite reasonable and obvious axioms LY added a comparison hypothesis (CH) saying that between two states $X$ and $Y$ at least one of these two transformations $X \prec Y$ or $Y \prec X$ is certain. At this level of our work we will consider $CH$ as an additional axiom.

\section{Ingredients for a thermodynamics of non-extensive systems}
In this Section we reconsider the axioms of LY in the case of non-extensive systems. The axioms $A_{1}$, $A_{2}$ and $A_{3}$ are very general and that they can be accepted in the case of non-extensive systems whatever the coordinate system considered.

\subsection{Axioms $A_{4}$}
Let consider an adiabatic transformation $X \prec Y$ that we can write 

\begin{equation}
{\bar{X}=(v_{X},\frac{U_{X}}{v_{X}},\frac{N_{X}}{v_{X}})\prec\bar{Y} = (v_{Y},\frac{U_{Y}}{v_{Y}},\frac{N_{Y}}{v_{Y}})}~ {\rm in~ICS}
\end{equation}

If dilations change $\bar{X}$ and $\bar{Y}$ into $a\bar{X}$ and $a\bar{Y}$ due to $H_{1}$ we have 

\begin{equation}
{\bar{X} \prec \bar{Y} \Rightarrow a\bar{X} \prec a\bar{Y}}
\end{equation}

By using $H_{2}$ we see that a common value of $a$ implies two different dilations in the $TCS$ we must introduce $t_{x}$ and $t_{y}$ such as 

\begin{equation}
a = t_{x}^{d(x)} = t_{y}^{d(y)}
\label{a}
\end{equation}

By using this relation we can write the new form of $A_{4}$ as 

\begin{equation}
X \prec Y \Rightarrow X(t_{x}) \prec Y(t_{y})
\label{a1}
\end{equation} 

If $(V, U, N)$ are the coordinates of $X$ those of $X(t_{x})$ are $(t_{x}V, t_{x}^{d(x)}U, t_{x}^{d(x)}N)$. The relation (\ref{a}) can be interpreted as follows. To associate a dilation to $X \prec Y$ we change the initial external volume from $V_{X}$ to $t_{x}V_{X}$ while the product of the transformation has to be analyzed in 
a volume $t_{y}V_{Y}$ and the relation between $t_{x}$ and $t_{y}$ is $t_{x}^{d(x)} = t_{y}^{d(y)}$; the value of $a$ in (\ref{a}) associated with the $ICS$ is an intermediate quantity. A generaliation of $A_{4}$ corresponds to   

\begin{equation}
(X_{1},\ldots, X_{n})\prec(Y_{1},\ldots, Y_{m})\Rightarrow (X_{1}(t_{x_{1}}),\ldots, X_{n}(t_{x_{n}}))\prec (Y_{1}(t_{y_{1}}),\ldots,Y_{m}(t_{y_{m}})) 
\label{an}
\end{equation} 

where the dilations verify

\begin{equation}
a = t_{x_{1}}^{d(x_{1})} = t_{x_{2}}^{d(x_{2})}  \cdots = t_{x_{n}}^{d(x_{n}} = t_{y_{1}}^{d(y_{1})} = t_{y_{2}}^{d(y_{2})} = \cdots = t_{y_{m}}^{d(y_{m})}
\label{nuple}
\end{equation}

Let apply $A_{4}$ to two arbitrary states $\bar{X'}$ and $\bar{Y'}$ defined in the $ICS$ and for which we have $\bar{X'} \prec \bar{Y'} \Rightarrow a\bar{X'} \prec a \bar{Y'}$, if $\bar{X'}$ and $\bar{Y'}$ are already the result of a dilation of the original states $\bar{X}$ and $\bar{Y}$ we have $b\bar{X} \prec b\bar{Y} \Rightarrow ab{X} \prec ab{Y}$. We are free to choose $ab = 1$ leading to

\begin{equation} 
b\bar{X} \prec b\bar{Y} \Rightarrow \bar{X} \prec \bar{Y}
\label{simpli}
\end{equation} 

\subsection{Axiom $A_{5}$}

In Section $3.2$ we have seen that the axiom $A_{5}$ corresponds to for any ${0<t<1}$, ${X\approx(tX,(1-t)X)}$. We have seen that such adiabatic equivalence seems obvious because splitting and recombinaition can be easily realized by using adiabatic walls. More generally for extensive systems we may write ${cX\approx(aX, bX)}$ provided ${a + b = c}$. A generalization 
of this axiom for non-extensive systems is less obvious because we cannot manipulate external adiabatic wall at the level of $v$. If the recombination can be easily performed the splitting must be considered as a theoretical process. Nevertheless we accept that this axiom subsists in the $ICS$ leading to $c \bar{X} \approx (a\bar{X}, b\bar{X})$ leading to three dilations $(t, t', t")$ in $TCS$ related according to 

\begin{equation}
t^{d(x)} = t'^{d(x)} + t''^{d(x)}
\label{t}
\end{equation} 

When (\ref{t}) is verified the new version of $A_{5}$ takes the form
\begin{equation}
X(t) \approx (X(t'), X(t''))
\label{a5}
\end{equation}

From the definition (\ref{sign}) if we have $X(t) \equiv ((X(t'), X(-t"))$ then (\ref{t}) is transformed into

\begin{equation}
t^{d(x)} = t'^{d(x)} - t''^{d(x)}
\label{a6}
\end{equation}

\subsection{Axiom $A_{6}$}  
Finally we slighlty generalize $A_{6}$. If $Z_{1}$ and $Z_{2}$ are two states of $\Gamma$ and if $f(\epsilon) \to 0$ when $\epsilon \to 0$ we accept that

\begin{equation}
(X,f(\epsilon)Z_{1}) \prec (Y,f(\epsilon)Z_{2})   \Rightarrow  X \prec Y
\label{a7}
\end{equation}

\subsection{Other conditions}
At this level the comparison hypothesis (CH) will be considered as an extra axiom. This hypothesis establishes that two states of $\Gamma$ are never independent since we must have, at least, $X \prec Y$ or $Y \prec X$.

It is convenient to add two extra obvious conditions to the definition of $\prec$. First, if $X \prec \prec Y$ we cannot find a given amount of $X$ and $Y$ for which the reverse transformation is possible. Second, we consider that the transformations $X \prec 0$ or that $0 \prec X$ are excluded they reject the possibility of anhilation or the production of material from nothing.

\subsection{Comparison with the Lieb Yngvason approach on non-extensive systems}
In a recent paper Lieb and Yngvason (\cite{liebnew}) have proposed an approach of non-extensive systems based on a generalization of their previous work. First, they define non-extensive systems as systems for which the axioms $A_{4}$ and $A_{5}$ do not exist. Second, they observe that it is difficult to work directly in a state space $\Gamma$ in which we drop these two Axioms. Due to this they introduce some states pertaining to a state space $\Gamma_{0}$ in which the axioms $A1 - A6$ and the comparability hypothesis hold accordingly the entropy is well defined in $\Gamma_{0}$. Their main idea is to form a couple of state $(X,Z)$ in which $X \in \Gamma$ and $Z \in \Gamma_{0}$, two references states are introduced one in $\Gamma$ and the other in $\Gamma_{0}$. By introducing a new axiom on the couple of states it is possible to associate the properties of $\Gamma$ to those of $\Gamma_{0}$ and the entropy of $X$ can be associated with the usual entropy of $Z$ calculated in $\Gamma_{0}$.

Here we keep the axioms $A_{4}$ and $A_{5}$ but we generalize them and we introduce two extra hypothesis. 
To given state we associate two coordinates system and for one, $ICS$, the extensivity is assumed. Then we have to deal with the following problematic: in the thermodynamic we must work with 
$TCS$ but the dilation process does not exist in the usual form, in the internal coordinate system $ICS$ the usual dilation is possible but the coordinates are not thermodynamic quantities. We can relate the $TCS$ to $ICS$ by introducing an extra law relating the extensive variables associated with each coordinate system. The strategy adopted is then the following. A given problem is first described in TCS where the external initial volume $V$ is under our control, an expansion of $V$ can be freely performed. Then we translate the problem in the $ICS$ where the demonstrations are very similar to the those developped in the original LY work and finally we return to the $TCS$. The knowledge of the scaling law for each state is crucial because the dilations that we have to perform may depend on the different states involved in the transformations of a state (see for instance (\ref{nuple})).

\section{Relations between two couples of states}
In this Section we establish two relations between two couples of states. The first gives the constraints concerning the possibility of transforming a couple and the second one is related to a stability condition.    
 
\subsection{Conditions for transformations of states}  
Let consider how to transform a couple of states $X_{0}$ and $X_{1}$ such as $X_{0} \prec \prec X_{1}$ in the same couple but with different extensions. This leads to investigate in $TCS$ the following transformation   

\ben 
\left(X_0(t_0),X_1(t_1)\right)\prec\left(X_0(t_0'), X_1(t_1')\right)
\label{prec1}
\een

that we can translate in $ICS$ for states such as $\bar{X}_{0}\prec\prec \bar{X}_{1}$ according to    

\ben 
\left(t_0^{d(x_0)}\bar{X}_0,t_1^{d(x_1)}\bar{X}_1)\right)\prec\left(t_0'^{d(x_0)}\bar{X}_0, t_1'^{d(x_1)}\bar{X}_1)\right)  
\label{prec2}
\een

or 

\begin{equation}
(t_{0}^{d(x_{0})}\bar{X}_{0}, - t_0'^{d(x_{0})}\bar{X}_{0})  \prec (t_1'^{d(x_{1})}\bar{X}_1, - t_1^{d(x_{1})}\bar{X}_{1})  
\label{prec3}
\end{equation}

By using the new version of $A_{5}$ we get

\begin{equation}
(t_{0}^{d(x_{0})} -t_0'^{d(x_{0})})\bar{X}_{0}  \prec (t_1'^{d(x_{1})} -t_{1}^{d(x_{1})})\bar{X}_{1}  
\label{prec4}
\end{equation}

First, note that we cannot have simultaneously 

\begin{equation}
(t_{0}^{d(x_{0})} -t_0'^{d(x_{0})}) < 0 ~{\rm and}~ (t_1'^{d(x_{1})} -t_{1}^{d(x_{1})}) <0
\end{equation}
because in this case (\ref{prec4}) can be rewritten 

\begin{equation}
(t_{1}^{d(x_{1})} -t_1'^{d(x_{1})})\bar{X}_{1} \prec (t_0'^{d(x_{0})} -t_{0}^{d(x_{0})}) \bar{X}_{0}
\label{prec5}
\end{equation}

showing that from a given amount of $\bar{X}_{1}$ it should be possible to reach a given amount of $\bar{X}_{0}$ in opposite with assumption of $\bar{X}_{0} \prec \prec \bar{X}_{1}$ (see subsection $IV.D$). Second, it is impossible for $(t_{0}^{d(x_{0})} -t_0'^{d(x_{0})})$ and $(t_1'^{d(x_{1})} -t_{1}^{d(x_{1})})$ to have opposite signs because to this case we will have anhilation or creation from nothing, two situations rejected in Section $IV. D$. Thus (\ref{prec4}) has a meaning if we have simultaneously 

\begin{equation}
(t_{0}^{d(x_{0})} -t_{0}'^{d(x_{0})}) > 0
\end{equation}

\begin{equation}
(t_1'^{d(x_{1})} -t_{1}^{d(x_{1})}) > 0. 
\label{positif}
\end{equation}

In order to have (\ref{prec4}) in agreement $\bar{X}_{0} \prec \bar{X}_{1}$ we must have from (\ref{simpli})  

\begin{equation}
(t_{0}^{d(x_{0})} -t_0'^{d(x_{0})}) = (t_1'^{d(x_{1})} -t_{1}^{d(x_{1})}).
\label{prec5} 
\end{equation}

Thus the transformation (\ref{prec1}) is compatible with $X_{0} \prec \prec X_{1}$ provided (\ref{prec5}) is verified with the two additional conditions of positivity (\ref{positif}). Of course, if $d(x_{0})= d(x_{1}) = 1$ we reobtain the results of LY.\\
It is easy to prove that (\ref{prec5}) and the positivity conditions implies (\ref{prec1}). By using $A_{5}$ and we obtain

\begin{equation}
(t_{0}^{d(x_{0})}\bar{X}_{0},t_{1}^{d(x_{1})}\bar{X}_{1}) \approx ((t_{0}^{d(x_{0})} - t_0'^{d(x_{0})})\bar{X}_{0},t_0'^{d(x_{0})}\bar{X}_{0},t_{1}^{d(x_{1})} X_{1}) 
\label {prec6}
\end{equation}

from $A_{5}$ and (\ref{prec4}) we deduce 

\begin{equation}
(t_{0}^{d(x_{0})}\bar{X}_{0},t_{1}^{d(x_{1})}\bar{X}_{1}) 
\prec ((t_1'^{d(x_{1})} - t_{1}^{d(x_{1})})\bar{X}_{1},t_0'^{d(x_{0})}\bar{X}_{0},t_{1}^{d(x_{1})} \bar{X}_{1})
\approx (t_0'^{d(x_{0})} \bar{X}_{0},t_1'^{d(x_{1})} \bar{X}_{1}) 
\label {prec7}
\end{equation}

\subsection{Stability condition}
It useful to demonstrate that for three states $X, Y, Z$ the transformation $(X, Z) \prec (Y, Z)$ implies $X \prec Y$. In order to prove this result we first proceed in the $ICS$ where we have to prove that $(\bar{X}, \bar{Z}) \prec (\bar{Y}, \bar{Z}) \Rightarrow \bar{X} \prec \bar{Y}$ showing that $\bar{X}$ can be transformed in $\bar{Y}$ independently of $\bar{Z}$. 
Let consider a small dilation of $Z$ leading to $Z(\epsilon)$ and taking $\epsilon \to 0$. In $ICS$ $\bar{Z}$ is transformed in $\epsilon^{d(z)}\bar{Z}$ and we put this state in contact with $\bar{X}$. The composed state $(\bar{X}, \epsilon^{d(z)}\bar{Z})$ is equivalent via $A_{5}$ to $ ((1 - \epsilon^{d(x)})\bar{X}, \epsilon^{d(x)}\bar{X}, \epsilon^{d(z)}\bar{Z})$. If we choose $t = \epsilon^{d(x)}$ we have now to deal with $((1-t)X, tX, tZ)$ and $t \to 0$ if $\epsilon \to 0$ since $d(x) > 0$. Using the same approach as in LY we find

\begin{equation}
(\bar{X} , \epsilon^{d(x)}\bar{Z}) \prec (\bar{Y} , \epsilon^{d(x)}\bar{Z})
\label{stability}
\end{equation}

and from the modified $A_{6}$ we conclude that 

\begin{equation}
(\bar{X}, \bar{Z}) \prec (\bar{Y}, \bar{Z}) \Rightarrow \bar{X} \prec \bar{Y} \Rightarrow X \prec Y.
\end{equation}

\section{Adiabatic equivalence}
To understand some choices introduced in the LY approach we first analyze the concept of adiabatic equivalence in standard thermodynamics.

\subsection{Adiabatic equivalence in standard thermodynamics}
Let consider two states $X$ and $X_{0}$ such that $X_{0} \prec X$ implying that their entropies $S(X_{0})$ and $S(X)$ are such as $S(X) \geq S(X_{0})$. Since the entropies are numbers and that the entropies are extensive we may perform a dilation $t$ of $X_{0}$ in order to have $S(tX_{0}) = tS(X_{0})= S(X)$. We can say that $X \approx  tX_{0}$ since these two states although different from a physical point of view have the same entropy and thus they can coexist in thermodynamic equilibrium. If $X_{0} \prec X$ can not be realized we may assume that we can find a state $X_{1}$ different from $X_{0}$ but from which we have $(X_{0}, X_{1}) \prec X$. Now using additivity and extensivity of the entropy in traditional thermodynamic we may find two numbers $t$ and $t'$ such as $S(X) = t'S(X_{0}) + tS(X_{1})$ leading to $X \approx (t'X_{0},tX_{1})$. To realize the entropy equality we have one equation but two parameters consequently one parameter can be eliminated by fixing an additional arbitrary condition, for instance $t + t' = 1$. Finally we can conclude that there is a value of $t$ for which 

\begin{equation}
X \approx ((1-t) X_{0}, t X_{1})
\label{eq1}
\end{equation}

All the ingredients that we have developped above appear in the LY approach but not in the same order and the non-equivalence of $X_{0}$ and $X_{1}$ is represented by $X_{0} \prec \prec X_{1}$. First, it is shown that for any state $X$ we can find a finite value of $t$ for which we have $((1-t)X_{0}, tX_{1}) \prec X$. In a second step we can show that it exists an 
extremum value for $t$ and finally it is possible to define a function verifying monotony, additivity and extensivity. Our work consists to extend this approach to the case of non-extensive systems. 

\subsection{Adiabatic equivalence in non-extensive systems}
Generalizing (\ref{eq1}) we want to show that it exists a couple of dilations $t$ and $t'$ such that

\begin{equation}
(\lambda X_{0}(t'), \mu X_{1}(t)) \prec X
\label{xtcs}
\end{equation}

or 

\begin{equation}
(\lambda t'^{d(x_{0})}\bar{X}_{0}, \mu t^{d(x_{1})}\bar{X}_{1 }) \prec \bar{X}
\label{xics}
\end{equation}

the two dilations being related by  

\begin{equation}
\mu t^{d(x_{1})} +  \lambda t'^{d(x_{0})} = 1
\label{relp}
\end{equation}

in which $\mu$ and $\lambda$ are two multiplicative factors taking the values $\pm 1$ chosen in order to verify (\ref{relp}). Thus when $t'^{d(x_{0})} > 1$ we  choose $\lambda = 1$ and $\mu = -1$ and the equation (\ref{xtcs}) becomes

\begin{equation}
(X_{0}(t'), -X_{1}(t)) \prec X   \Rightarrow  X_{0}(t') \prec (X,X_{1}(t))
\label{tnegat}
\end{equation}

with 

\begin{equation}
t'^{d(x_{0})} = 1 + t^{d(x_{1})}  
\label{relpnegat}
\end{equation} 

Now we follow the same route as in LY consisting to show that if (\ref{xics}) is not verified we are in contradiction with our starting point $X_{0} \prec \prec X_{1}$. For very large values of $t'$ for which we have $t'^{d(x_{0})} > 1$ we have to use (\ref{tnegat}) and (\ref{relpnegat}).  from (\ref{relpnegat}) we see that very large values of $t'$ implies very large values of $t$. If (\ref{tnegat}) is not verified from CH we have to verify

\begin{equation}
(\bar{X}, t^{d(x_{1})}\bar{X}_{1}) \prec t'^{d(x_{0})}\bar{X}_{0}
\label{negat1}
\end{equation}      

that we can rewrite as 

\begin{equation}
t^{d(x_{1})} (\bar{X}_{1},\frac{1}{t^{d(x_{1})}}\bar{X}) \prec t^{d(x_{1})}\bar{X}_{0}( 1 + \frac{1}{t^{d(x_{1})}})  
\approx t^{d(x_{1})}(\bar{X}_{0}, \frac{1}{t^{d(x_{1})}}\bar{X}_{0})
\label{negat2}
\end{equation}

where (\ref{relpnegat}) and $A_{5}$ have been used. 
After a cancellation by $t^{d(x_{1})}$ (see (\ref{simpli})) we obtain   

\begin{equation}
(\bar{X}_{1}, \frac{1}{t^{d(x_{1})}}\bar{X}) \prec  (\bar{X}_{0},\frac{1}{t^{d(x_{1})}} \bar{X}_{0})
\label{negat3}
\end{equation} 

now if we take $t = \frac{1}{\epsilon}$ and $\epsilon \to 0$ we may use the modified version of $A_{6}$ showing that $\bar{X}_{1} \prec \bar{X}_{0}$ or ${X}_{1} \prec {X}_{0}$ in contradistinction with our initial assumption. We can conclude that, in general, (\ref{negat1}) is wrong and thus (\ref{xtcs}) and (\ref{xics}) are verified. \\
Using the same kind of demonstration it is possible to prove that $t$ and then $t'$ are finite. If $t^{d(x_{1})} > 1$ we choose $\mu = 1$ and $\lambda = -1$ from (\ref{relp}) we have

\begin{equation}
(-X_{0}(t'), X_{1}(t)) \prec X  \Rightarrow (\bar{X}_{1}, \frac{1}{t^{d(x_{1})}}\bar{X}_{0}) \prec (\bar{X}_{0}, \frac{1}{t^{d(x_{1})}}\bar{X})
\label{agrand}
\end{equation} 

If we introduce  $\epsilon \to 0$ and $t = \frac{1}{\epsilon} \to \infty$ and use $A_{6}$ from $\ref{agrand}$ we obtain $ X_{1} \prec X_{0}$ that contradicts our hypothesis and we can conclude that $t$ must be finite and thus $t$ has a maximum finite value $\bar{t}$ while $\bar{t'}$ the corresponding value of $t'$ is given by (\ref{relp}) and (\ref{xtcs}) is verified for these values of $t$ and $t'$.

\subsection{Optimization}
Since $\bar{t}$ is an external parameter we are free to consider the dilations $\bar{t} + \epsilon_{1}$ and $\bar{t'} + \epsilon_{0}$. We consider the case $\lambda = \mu = 1$ generalizations are obvious. For $\epsilon_{1} > 0$ the transformation 
(\ref{xics}) can not be verified because we are beyond the maximum and then due to CH we have 

\begin{equation}
\bar{X} \prec ((1-(\bar{t} + \epsilon_{1})^{d(x_{1})})\bar{X}_{0},(\bar{t} + \epsilon_{1})^{d(x_{1})}\bar{X}_{1})
\label{max1}
\end{equation}

For vanishingly small values of $\epsilon_{1}$ we have $(\bar{t} + \epsilon_{1})^{d(x_{1})} \approx 
\bar{t}^{d(x_{1})} + \epsilon_{1}d(x_{1}) \bar{t}^{(d(x_{1}) -1)}$ and by using $A_{5}$ we obtain

\begin{equation}
(\bar{X}, \epsilon_{1}d(x_{1})\bar{t}^{(d(x_{1}) -1)} \bar{X}_{0}) \prec ((1 - \bar{t}^{d(x_{1})})\bar{X}_{0}, \bar{t}^{d(x_{1})}\bar{X}_{1}, \epsilon_{1}d(x_{1})\bar{t}^{(d(x_{1}) -1)} \bar{X}_{1})
\label{max2}
\end{equation}

and from $A_{6}$ we get $X \prec (X_{0}(\bar{t'}),X_{1}(\bar{t}))$. For a dilation $\bar{t} - \epsilon_{1}$ the transformation $(\ref{xtcs})$ is verified and by using (\ref{relp}), the expansion in terms of $\epsilon$, $A_{5}$ and $A_{6}$ it is easy to prove that $((X_{0}(\bar{t'}),X_{1}(\bar{t}) \prec X$ and in addition with the previous result $(X_{0}(\bar{t'}),X_{1}(\bar{t}) \prec \bar{X})$ we have established that 

\begin{equation}
X \approx (X_{0}(\bar{t'}),X_{1}(\bar{t}))
\label{equivalence}
\end{equation} 

in which $t'$ and $t$ are related by (\ref{relp}).

It is interesting to investigate what happens if we perform a dilation starting from $\bar{X} \approx (\bar{t'}^{d(x_{0})}\bar{X}_{0}, \bar{t}^{d(x_{1})}\bar{X}_{1})$ in the case $\lambda = \mu = 1$ (a generalization is trivial) and for which we have $\bar{t'}^{d(x_{0})} + \bar{t}^{d(x_{1})}$. A dilation $t"$ of $X$ leads to  

\begin{equation}
t"^{d(x)}\bar{X} \approx (t"^{d(x)}\bar{t'}^{d(x_{0})}\bar{X}_{0}, t"^{d(x)}\bar{t}^{d(x_{1})}\bar{X}_{1})
\label{max}
\end{equation}

The state $t"^{d(x)}\bar{X}$ can be also expressed on the same reference states $\bar{X}_{0}$ and $\bar{X}_{1}$ according to $t"^{d(x)}\bar{X} \approx (\bar{u'}^{d(x_{0})}\bar{X}_{0}, \bar{u}^{d(x_{1})}\bar{X}_{1})$ in which $\bar{u}^{d(x_{1})}$ is an extremum for this quantity; we have $\bar{u'}^{d(x_{0})} + \bar{u}^{d(x_{1})} = t"^{d(x)}$. If $\bar{u}^{d(x_{1})}$ does 
not correspond to $t"^{d(x)}\bar{t}^{d(x_{1})}$ we write  $\bar{u}^{d(x_{1})} = t"^{d(x)}\bar{t}^{d(x_{1})} + f_{1}t"^{d(x)}$  and 
$\bar{u'}^{d(x_{0})} = t"^{d(x)}\bar{t'}^{d(x_{0})} + f_{0}t"^{d(x)}$. The relation between $\bar{u}$ and $\bar{u'}$ leads to $t"^{d(x)}\bar{t'}^{d(x_{0})} + f_{0}t"^{d(x)} + t"^{d(x)}\bar{t}^{d(x_{1})} + f_{1}t"^{d(x)} = t"^{d(x)}$ and finally to $f_{1} + f_{0} = 0$. When we bring all these results together we get 

\begin{equation}
\bar{X} \approx (\bar{X}, -f_{1}\bar{X}_{0}, f_{1}\bar{X}_{1})
\label{max}
\end{equation}

If $f_{1}$ does not vanish in order to verify (\ref{max}) we see that $X_{0}$ must be equivalent to $X_{1}$ in opposite to our hypothesis. Tus we must have $f_{1} = 0$ proving that

\begin{equation}
\bar{u}^{d(x_{1})} = t"^{d(x)} \bar{t}^{d(x_{1})}
\label{maxa}
\end{equation} 

Thus, if we know the value of $\bar{t}^{d(x_{1})}$ after a dilation $t"$ of this state this quantity is replaced by $\bar{u}^{d(x_{1})}$ according (\ref{maxa}). 

\section{Entropy}
The adiabatic equivalence established in the previous Section will be used to introduce the entropy; to save the notations the optimum values $\bar{t}$ will be replaced by $t$ in this Section and we consider the particular case $\lambda=\mu=1$ (generalization are straightforward). We first consider the transformation $X\prec Y$ and from the equivalences $X \approx (X_{0}(t_{0}),X_{1}(t_{1}))$ and $Y \approx (X_{0}(t_0'),X_{1}(t_1'))$ this transformation can be written in the form (\ref{prec1}), the constraint (\ref{prec5}) is automatically verified because $X$ and $Y$ verify separately (\ref{relp}) and their combination gives (\ref{prec5}) and from (\ref{positif}) we must have $(t_1'^{d(x_{1})} > t_{1}^{d(x_{1})})$.
If the transformation is reversible we have also $Y \prec X$ leading to $(t_{1}^{d(x_{1})} > t_1'^{d(x_{1})}) > 0$; thus to conciliate the two inequalities we must have $t_1'^{d(x_{1})} = t_{1}^{d(x_1)}$ in the case of  a reversible transformation. If the two states $\bar{X}$ and $\bar{Y}$ can be represented as two multicomponent states $\bar{X} = (\bar{X}_{1}, \bar{X}_{2},\ldots, \bar{X}_{n}) $ and $\bar{Y} = (\bar{Y}_{1}, \bar{Y}_{2},\ldots,\bar{Y}_{m})$ the adiabatic equivalence leads to transform $\bar{X} \prec \bar{Y}$ in $ICS$ according to

\begin{equation}
(\bar{X}_{0}\sum^{n}_{1}t_{0i}^{d(x_{0})}, \bar{X}_{1}\sum^{n}_{1}t_{1i}^{d(x_{1})})\prec (\bar{X}_{0}\sum^{m}_{1}t_{0j}'^{d(x_{0})}, \bar{X}_{1}\sum^{m}_{1}t_{1j}'^{d(x_{1})})
\label{ent1}
\end{equation}

in which $t_{0i}$ and $t_{1i}$ are the dilation associated with the state $X_{i}$. From (\ref{positif}) we conclude that

\begin{equation}
\bar{X}\prec\bar{Y} \Rightarrow \sum^{m}_{1}t_{1j}'^{d(x_{1})} \geq \sum^{n}_{1}t_{1i}^{d(x_{1})}
\label{ent2}
\end{equation}     

If we define $S(X)=\sum^{n}_{1}t_{1i}^{d(x_1)}$ and $S(Y)=\sum^{n}_{1}t_{1i}'^{d(x_1)}$ we see that this quantity verifies the monotony of the thermodynamic entropy since $X \prec Y \Rightarrow S(Y) \geq S(X)$. If we introduce $S(X_{i}) = t_{1i}^{d(x_1)}$ as the entropy of the state $X_{i}$ we have $S(X)=\sum^{n}_{i=1}S(X_i)$ showing that $S(X)$ is an additive quantity. By construction we have $t_{1i} > 0$ and then $S(X_i) > 0$ and $S(X)> 0$. By using (\ref{maxa}) if $\alpha_{i}$ is a dilation of the state $i$ we have the following scaling law 

\begin{equation}
S(X_{i}(\alpha_{i})) = \alpha_{i}^{d(x_{i})}S(X_{i})
\label{scaling}
\end{equation} 

showing that the entropy is no more an extensive quantity.

We can verify that our definition of $S(X)$ is consistent with $X\approx(X_{0}(t_0),X_{1}(t_1))$. By the additivity we have 
$S(X)= S(X(t_{0})) +S(X(t_{1}))$ and from the scaling law (\ref{scaling}) we have 
$S(X)=t_{0}^{d(x_0)} S(X_0) + t_{1}^{d(x_1)} S(X_1)$.
When we use $X_{0}$ and $X_{1}$ as reference states it is easy to verify that $S(X_{1}) =1$ and $S(X_{0}) =0$. For instance, we have $X_{1} \equiv ((X_{0}(u_{0}),X_{1}(u_{1})))$ with $u_{0}^{d(x_{0})} + u_{1}^{d(x_{1})} = 1$. Since with the reference states we do not need $X_{0}$ to reach $X_{1}$ we can take $u_{0} = 0$ and then $u_{1} = 1$ which is the optimum value of 

$u_{1}^{d(x_1)}$. Thus on these reference states we have 

$S(X_1)=1$ and using similar arguments we can obtain that 

$S(X_0)=0$ showing that 

$X\approx(X_{0}(t_{0}),X_{1}(t_{1}))$ is consistent with our choice for the entropy 
${S(X)=t_{1}^{d(x_{1})}}$.

We now consider what happens if we change the reference states $X_{0}$ and $X_{1}$ for which the corresponding entropy can be noted $S(X)$ to new reference states $X'_{0}$ and $X'_{1}$ where the entropy is $S'(X)$. By writing that $X_{0}$ and $X_{1}$ have adiabatic equivalents in terms of $X'_{0}$ and $X'_{1}$ and tacking into account that the dilations must verfy (\ref{relp}) straightforward calculations lead to

\begin{equation}
S'(X) = S(X)(S'(X_{1}) - S'(X_{0})) + S'(X_{0})
\label{end}
\end{equation}

Since $X_{0} \prec X_{1}$ whatever the reference system we have $S'(X_{1}) - S'(X_{0}) > 0$ and $S'(X) > 0$ as expected. The relation (\ref{end}) shows that a change of the reference states leads to an affine transformation of the entropy.   

\section{Conclusion}
For traditional mono-component system we know from the Gibbs phase rule that a system is determined by one extensive property - here we have chosen the volume $V$ in which the system is inclosed - and two intensive variables. The other extensive properties are proportional to $V$. In particular this is the case of the thermodynamic entropy. However it exists a large class of real systems for which the linearity with $V$ is lost. In Section $2$ we have characterized these non-extensive systems by two assumptions $H_{1}$ and $H_{2}$ induced from experimental facts. With $H_{1}$ we assume that, in parallel with the $TCS$ it exists and internal coordinate system $ICS$ in which the extensivity is verified. Then we have to deal with the following dilemma: on one side it exists a thermodynamic coordinates system, $TCS$, which corresponds to the physical parameters that are under our control but on the $TCS$ we are not able to characterize the non-extensivity and on another side we have extensivity on a internal coordinates system $ICS$ for which the volume is not diretly under our control. This internal volume that we can determine by experiments is not a true thermodynamic quantity because it is related to a molecular system description. The relation between $ICS$ and $TCS$ is assured via an extra relation between an extensive variable defined both in $TCS$ and $ICS$ this is our hypothesis $H_{2}$. To be illustrative we have assumed the existence of scaling law related the volumes in the two coordinates systems. The difference between $TCS$ and $ICS$ can be illustrated in the case of a dilation process, in $ICS$ an expansion of the internal volume needs an identical expansion for the energy and the number of particles while a linear expansion of the external volume $V \rightarrow tV$ will require in $TCS$ an expansion of the internal energy and number of particles related to the scaling law. However we will assume that a physical transformation exists in $TCS$ as well as in $ICS$. Due to this we may repeat in $ICS$ the mathematical processes developped in LY. But when we return to the $TCS$ physical differences exist.  
This can be illustrated with the transformation $X \prec Y$. In extensive systems it is assumed that such transformation is possible whatever the amount of material and thus $\forall t > 0$ we have $X \prec Y \Rightarrow tX \prec tY$. For non-extensive systems the dilation persists but between states noted $X(t_{x})$ and $Y(t_{y})$ for which $t_{x} \neq t_{y}$ and their relation depends on the state of the system. Our main result is the following: for non extensive systems it exists a function veryfying the  monotony and the additivity of the usual thermodynamic entropy but the extensivity of the entropy is lost. Nevertheless we may consider that the main properties of the entropy survive and we conclude that non-extensive systems verify the second law of thermodynamics.

\end{document}